\documentclass[preprint,aps]{revtex4}

\usepackage{graphicx}
\usepackage{amsmath}

\begin{document}
\draft
\title{A feasible gate for scalable linear optics quantum computation using polarization optics}
\author{Kaoru Sanaka and Kevin Resch}
\address{Institut f$\ddot{u}$r Experimentalphysik, Universit$\ddot{a}$t Wien, Boltzmanngasse\\
5, A-1090 Wien, Austria}
\date{\today}

\begin{abstract}
Knill, Laflamme, and Milburn (KLM) proved that it is possible to
build a scalable universal quantum computer using only
linear-optics elements and conditional dynamics [Nature (London)
{\bf 409}, 46 (2001)\cite{Knill}]. However, the practical
realization of the quantum logic gate for the scheme is still
technically difficult. A major difficulty is the requirement for
sub-wavelength level stabilization of the interlocking
interferometers. Following our recent experimental work[Phys.
Rev. Lett.{\bf 92}, 017902 (2004)\cite{Sanaka2}], we describe a
more feasible scheme to implement the gate that greatly reduces
the experimental stability requirements. The scheme uses only
polarizing beam splitters and half-wave plates.
\end{abstract}

\maketitle

\preprint{HEP/123-qed}

\narrowtext

Quantum computers promise unprecedented processing power for
certain types of problems like factoring products of large prime
numbers\cite{Shor} and searching database \cite{Grover}.We have
recently experimentally demonstrated many important features of
the nonlinear sign shift (NS) operation in a stable, polarization
insensitive configuration for the scalable linear optics quantum
computation\cite{Sanaka2}. In this theoretical work, we show how
those NS operations can be used, together with additional linear
optical elements, to build up a two qubit logic gate.

One of the most promising systems in which to store and manipulate
quantum information is the polarization state or spatial mode of
single photons. This is due to the photon's robustness against
decoherence and the ease of which certain necessary operations
(i.e., single-qubit operations) can be carried out. However it
has been very difficult to perform the necessary two-qubit
operations since there is no natural interaction between photons;
even when atoms mediate photon-photon interactions through
nonlinear optical effects, these effects are usually negligible
at the quantum level. Nonlinear optics at the quantum level has
been demonstrated in cavity QED
experiments\cite{Turchette,Rauschenbeutel}, through quantum
interference\cite{Resch}, and proposed for systems using
electromagnetically-induced transparency\cite{Harris,Kash} and
photon-exchange interactions\cite{Franson}. In stark contrast to
these approaches, Knill, LaFlamme, and Milburn (KLM) showed that
effective nonlinear interactions could be implemented using only
linear optical elements in conjunction with single-photon sources
and conditioning evolution based on successful
measurements\cite{Knill}. Furthermore, they proved that their
scheme could be used, in principle, to build a scalable universal
quantum computer. While their two-qubit gate itself is not
deterministic, KLM showed that it can be made arbitrarily close
to deterministic when used in conjunction with quantum
teleportation \cite{Gottesman}.

In the KLM scheme, the two-qubit gate is achieved by the
conditional sign-flip(CS) operation : $|M\rangle
_{\text{A}}|N\rangle _{\text{B}}\rightarrow \ exp(\pi
MN)\,|M\rangle _{\text{A}}|N\rangle _{\text{B}}\,$. Where
$M=\{0,1\}$, $N=\{0,1\}$ show the number of photons in optical
mode A and B. The CS gate is constructed by the most fundamental
element, so-called nonlinear sign-shift (NS) gates, which
performs the transformation: $\alpha |0\rangle+ \beta |1\rangle
+\gamma |2 \rangle \rightarrow \alpha |0\rangle+ \beta
|1\rangle-\gamma |2 \rangle $ with some probability. Where the
kets represents the number of photons in a optical path. CS
operation is achieved by placing one NS gate in each arm of a
balanced Mach Zehnder interferometer.

Some other types of CS that are not based on the KLM NS gate have
also been proposed
\cite{Knill2,Koashi,Pittman,Zou,Hofmann,Sanaka}, however Lund
{\em et al.} shows that the two-qubit gate using the NS gates is
the most resilient with non-ideal ancilla mode production and
detection\cite{Lund}. Ralph {\em et al.} simplified the KLM NS
gate using non-polarizing beam splitters with unequal
reflectivities\cite{Ralph}. This considerable flexibility makes
it possible to construct the CS gate with much lower effort than
original KLM scheme. However the sub-wavelength level
stabilization for the whole setup is still essential because the
injected photons are spatially superposed with the two arms of
Mach Zehnder interferometer and the phase correlation between the
arms must be maintained during the operations.The sensitivity on
the optical phase makes the practical realization of their CS
gate extremely difficult. Here we show that a significant
simplification is possible to implement the CS gate by using the
different polarization states in a same spatial mode by removing
or strongly reducing the phase sensitivity of the system.

\begin{figure}[tbp]
\includegraphics{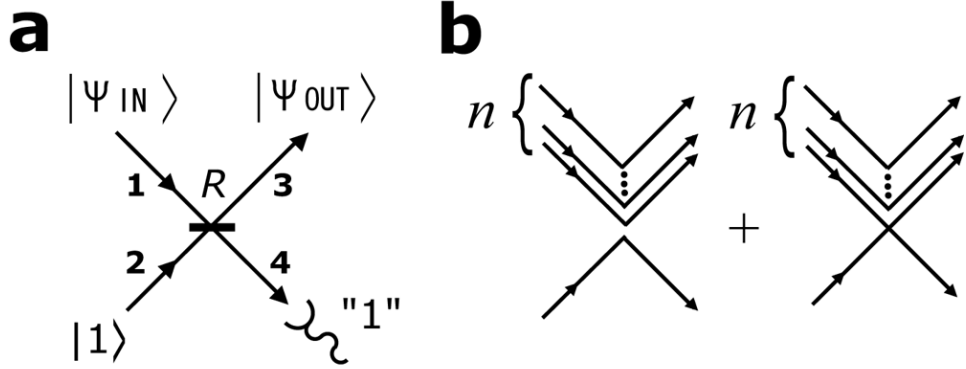}
\caption{(A) Schematic of a simplified version of nonlinear
sign-shift (NS) operation constructed by a non-polarizing beam
splitter of reflectivity $R$. $|\Psi _{\text{IN}}\rangle $ and
$|\Psi _{\text{OUT}}\rangle $ are the quantum states of input and
output photons. The operation is successful when the
single-photon detector in ancilla mode 4 counts a single photon.
(B) The two paths that lead to the detection of exactly one
photon in output mode 4. In the first case all of the photons are
reflected. In the second case, one ancilla photon and one of the
n input photons are transmitted and rest are reflected. As long
as all of the photons are indistinguishable, these two paths can
interfere.} \label{fig1}
\end{figure}

A simplified version of NS operation is shown in
Fig.\ref{fig1}(a). An input state, $|\Psi _{\text{IN}}\rangle
=|n\rangle $, impinges on a beam-splitter (BS) with reflection
probability $R$; a single ancilla photon, $|1\rangle \,$impinges
from the other side of the beam splitter. The two input modes, 1
and 2, undergo a unitary transformation into
two output modes, 3 and 4, described by $\widehat{a}_{1}\rightarrow \sqrt{R}\,\widehat{a%
}_{3}+\sqrt{1-R}\,\widehat{a}_{4}\,$and $\widehat{a}_{2}\rightarrow -\sqrt{%
1-R}\,\widehat{a}_{3}+\sqrt{R}\,\widehat{a}_{4}\,$. The NS
operation is successful when one and only one photon reaches the
detector in mode 4. Provided the photons are indistinguishable,
the two paths leading to exactly one photon in mode 4 will
interfere. Either all $n+1$ photons are reflected or $n-1$ of the
photons in mode 1 are reflected and 1 photon in each of modes 1
and 2 are transmitted. When a single photon ends up in mode 4, the
photon number state undergoes the following transformation:
\begin{equation}
|\Psi _{\text{IN}}\rangle =|n\rangle
\stackrel{\text{NS}}{\rightarrow }|\Psi _{\text{OUT}}\rangle
=(\sqrt{R})^{n-1}[R-n(1-R)]\,|n\rangle ,  \label{equ1}
\end{equation}
where the unusual normalization of the output state reflects the
probability amplitude of success. The sign of the phase shift
depends on the number of incident photons and the reflection
probability of the BS. For $n<R/(1-R)$, the sign of the amplitude
is unchanged and for $n>R/(1-R)$ it picks up a negative sign. For
$n=R/(1-R)$ the critical case, where , the output probability
amplitude becomes zero. This is due to the generalized
Hong-Ou-Mandel-style interference effect\cite{Hong}. In our
proposal, the extension of the NS operation to include a second
polarization mode is straightforward. We inject a
horizontally-polarized ancilla photon into the BS in Fig.
\ref{fig1} and consider only the cases where the single photon
detected in mode 4 is horizontally polarized. The transformation
for the horizontal polarization is as in (\ref{equ1}). There is
only 1 possible path which leads to no vertically-polarized
photons in mode 4; that is for all vertically-polarized photons
to be reflected. This operation for the input state with $m$
vertically-polarized photons and $n$ horizontally-polarized
photons is given by:
\begin{eqnarray}
|\Psi _{\text{IN}}\rangle &=&|m_{\text{V}};n_{\text{H}}\rangle  \nonumber \\
\stackrel{\text{NS}}{\rightarrow }|\Psi _{\text{OUT}}\rangle &=&(\sqrt{R_{%
\text{V}}})^{m}(\sqrt{R_{\text{H}}})^{n-1}[R_{\text{H}}-n_{\text{H}}
(1-R_{\text{H}})]\,|m_{\text{V}};n_{\text{H}}\rangle ,
\label{equ2}
\end{eqnarray}
where $R_{\text{V}}$ and $R_{\text{H}}$ are the reflection
probabilities for vertical and horizontal polarization
respectively. As expected, the vertical photon number, $m$, does
not appear in the square bracket nonlinear-sign term. The only
change the vertically-polarized photons contribute is the
reflection amplitude raised to the power of $m$.

\begin{figure}[tbp]
\includegraphics{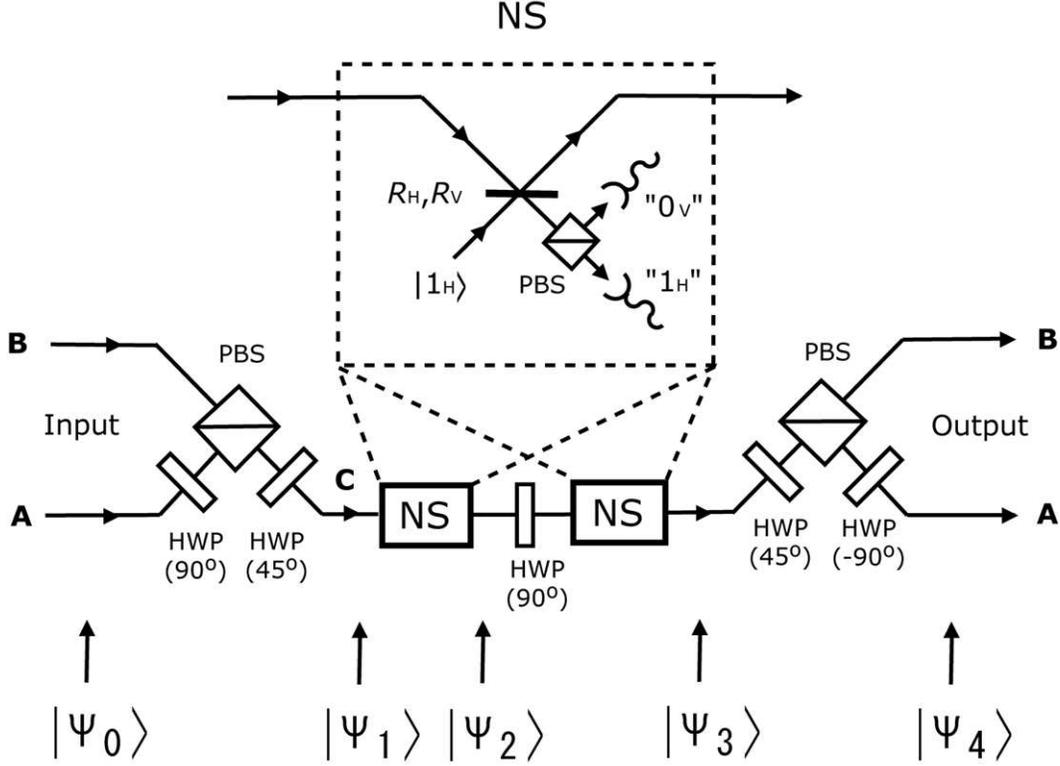}
\caption{Two-qubit gate for the full conditional-sign flip which
implements the transformation (for input states 00, 01, 10, 11)
using two polarization-based NS gates and linear-optics elements.
Two incoming horizontally-polarized modes are combined into a
single spatial mode (with different polarizations) using
half-wave plates (HWP), and a polarizing beam splitter (PBS).
(The brackets of HWP show the rotation angle.) The NS operations
are performed twice, once each for each polarization, and the
photons are reconverted back to two separate spatial modes. The
inset shows the detail of polarization-based NS gate. It is
constructed using a beam splitter with polarization-sensitive
reflectivity probabilities $R_{\text{V}} $ and $R_{\text{H}}$ and
single-photon detectors. Single-photon sources,
$|1_{\text{H}}\rangle $ are injected into one input port of BS's
as ancilla. The successful operation is signaled when the single
photon detector monitoring the horizontal mode in each NS
operation detects exactly one photon and no vertically-polarized
photons are detected.} \label{fig2}
\end{figure}

The operation of CS gate is constructed on the setup using two
such polarization-based NS gates and linear-optics element shown
in Fig. \ref{fig2}. The inset shows the detail of the
polarization-based NS gate. It is constructed using a beam
splitter with polarization-sensitive reflectivity probabilities
$R_{\text{V}}$ and $R_{\text{H}}$ and single-photon detectors.
Single-photon sources, $|1_{\text{H}}\rangle $ are injected into
one input port of BS's as ancilla. The successful operation is
signaled when the single photon detector monitoring the
horizontal mode in each NS operation detects exactly one photon
and no vertically-polarized photons are detected using a
polarizing beam-splitter (PBS) and single-photon detectors.

We consider input states $|\Psi _{\text{0}}\rangle $ that are
superpositions
of $\,|0\rangle _{\text{A}}\,|0\rangle _{\text{B}}$, $|0\rangle _{\text{A}%
}|1_{\text{H}}\rangle _{\text{B}}$, $|1_{\text{H}}\rangle _{\text{A}%
}|0\rangle _{\text{B}}$ and $|1_{\text{H}}\rangle _{\text{A}}|1_{\text{H}%
}\rangle _{\text{B}}$ ,where all of the photons are horizontally
polarized. This forms a computational space for 2 qubits. The
general input state for the gate is described as:
\begin{equation}
|\Psi _{\text{0}}\rangle =a\,|0\rangle _{\text{A}}\,|0\rangle _{\text{B}%
}+b\,|0\rangle _{\text{A}}|1_{\text{H}}\rangle _{\text{B}}+c\,|1_{\text{H}%
}\rangle _{\text{A}}|0\rangle _{\text{B}}+d\,|1_{\text{H}}\rangle _{\text{A}%
}|1_{\text{H}}\rangle _{\text{B}}.  \label{psi_0}
\end{equation}
The two input optical modes are combined into a single spatial
mode using a HWP that rotates the polarizations by 90 degrees and
a PBS. Once in the same spatial mode, second HWP rotates the
polarizations by 45 degrees. Following the transformation by the
setup, the combined state in mode C is described as:
\begin{eqnarray}
|\Psi _{\text{1}}\rangle  &=&a\,|0_{\text{V}};0_{\text{H}}\rangle _{\text{C}%
}+\frac{b-c}{\sqrt{2}}|0_{\text{V}};1_{\text{H}}\rangle _{\text{C}}+\frac{b+c%
}{\sqrt{2}}|1_{\text{V}};0_{\text{H}}\rangle _{\text{C}}  \nonumber \\
&&-\frac{d}{\sqrt{2}}|0_{\text{V}};2_{\text{H}}\rangle _{\text{C}}+\frac{d}{%
\sqrt{2}}|2_{\text{V}};0_{\text{H}}\rangle _{\text{C}}.
\label{psi_1}
\end{eqnarray}
The first NS operation is applied for $|\Psi _{\text{1}}\rangle $
and affects only the horizontal component of the state. Following
the polarization based NS operation (\ref{equ2}), the state in
mode C is undergoes the following transformation:
\begin{eqnarray}
|\Psi _{\text{2}}\rangle  &=&\sqrt{R_{\text{H}}}a\,|0_{\text{V}};0_{\text{H}%
}\rangle _{\text{C}}  \nonumber \\
&&-\frac{(1-2R_{\text{H}})(b-c)}{\sqrt{2}}|0_{\text{V}};1_{\text{H}}\rangle
_{\text{C}}+\frac{\sqrt{R_{\text{V}}R_{\text{H}}}(b+c)}{\sqrt{2}}|1_{\text{V}%
};0_{\text{H}}\rangle _{\text{C}}  \nonumber \\
&&+\frac{\sqrt{R_{\text{H}}}(2-3R_{\text{H}})\,d}{\sqrt{2}}|0_{\text{V}};2_{%
\text{H}}\rangle _{\text{C}}+\frac{R_{\text{V}}\sqrt{R_{\text{H}}}\,d}{\sqrt{%
2}}|2_{\text{V}};0_{\text{H}}\rangle _{\text{C}}.  \label{psi_2}
\end{eqnarray}
A HWP that rotates the polarizations by 90 degrees after the
first NS gate exchanges the roles of horizontal and vertical
polarization. Then, the second NS operation is applied and
affects the horizontal component of the state.
\begin{eqnarray}
|\Psi _{\text{3}}\rangle  &=&R_{\text{H}}\,a\,|0_{\text{V}};0_{\text{H}%
}\rangle _{\text{C}}  \nonumber \\
&&+\frac{\sqrt{R_{\text{V}}R_{\text{H}}}(1-2R_{\text{H}})}{\sqrt{2}}%
\,\{(b+c)\,|0_{\text{V}};1_{\text{H}}\rangle _{\text{C}}-(b-c)\,|1_{\text{V}%
};0_{\text{H}}\rangle _{\text{C}}\}  \nonumber \\
&&-\frac{R_{\text{V}}R_{\text{H}}\,(2-3R_{\text{H}})\,d}{\sqrt{2}}\,\,(|0_{%
\text{V}};2_{\text{H}}\rangle
_{\text{C}}-|2_{\text{V}};0_{\text{H}}\rangle _{\text{C}}).
\label{psi_3}
\end{eqnarray}
Using the final HWP rotates the polarizations by 45 degrees and a
polarizing beam-splitter (PBS), and HWP rotates the polarizations
by -90 degrees combination, the photons are split up again into
two different spatial modes and returned to the same polarization
states.
\begin{eqnarray}
|\Psi _{\text{4}}\rangle  &=&R_{\text{H}}\,a\,\,|0\rangle _{\text{A}%
}\,|0\rangle _{\text{B}}  \nonumber \\
&&+\sqrt{R_{\text{V}}R_{\text{H}}}(1-2R_{\text{H}})\,\,(b\,\,|0\rangle _{%
\text{A}}|1_{\text{H}}\rangle _{\text{B}}+c\,\,|1_{\text{H}}\rangle _{\text{A%
}}|0\rangle _{\text{B}})  \nonumber \\
&&-R_{\text{H}}R_{\text{V\thinspace }}(2-3R_{\text{H}})\,d\,|1_{\text{H}%
}\rangle _{\text{A}}|1_{\text{H}}\rangle _{\text{B}}.
\label{equ4}
\end{eqnarray}
When the BS has reflection probabilities:{\setcounter{enumi}{%
\value{equation}} \addtocounter{enumi}{1} \setcounter{equation}{0} %
\renewcommand{\theequation}{\theenumi\alph{equation}}
\begin{eqnarray}
R_{\text{V}} &=&5-3\sqrt{2}\approx 0.76,  \label{rv} \\
R_{\text{H}} &=&(3-\sqrt{2})/7\approx 0.23,  \label{rh}
\end{eqnarray}
\setcounter{equation}{\value{enumi}} }then achieving
$R_{\text{H}}\,=\sqrt{R_{\text{V}}R_{\text{H}}}(1-2R_{\text{H}})=R_{\text{H}}R_{\text{V\thinspace
}}(2-3R_{\text{H}})$ and the two-qubit CS gate is implemented
because all probability amplitudes of the output states for
$\,|0\rangle _{\text{A}}\,|0\rangle _{\text{B}}$, $|0\rangle
_{\text{A}}|1_{\text{H}}\rangle _{\text{B}}$,
$|1_{\text{H}}\rangle _{\text{A}}|0\rangle _{\text{B}}$ and
$|1_{\text{H}}\rangle _{\text{A}}|1_{\text{H}}\rangle
_{\text{B}}$  become equal. The required reflectivities are the
same as those of non-polarizing beam splitters for the Ralph's
gate and both of the gates give the same probability of successful
operation\cite{Ralph}. The success probability is given by
$R_{\text{H}}^{2}\approx 0.05$. This two-qubit gate is
non-deterministic, however near deterministic operation is
possible when the gate is used in conjunction with a quantum
teleportation protocol \cite{Knill,Gottesman}. Universal quantum
computation is then possible with this two-qubit gate together
with all single-qubit rotations from a universal set
\cite{Sleator,Barenco}.

\begin{figure}[tbp]
\includegraphics{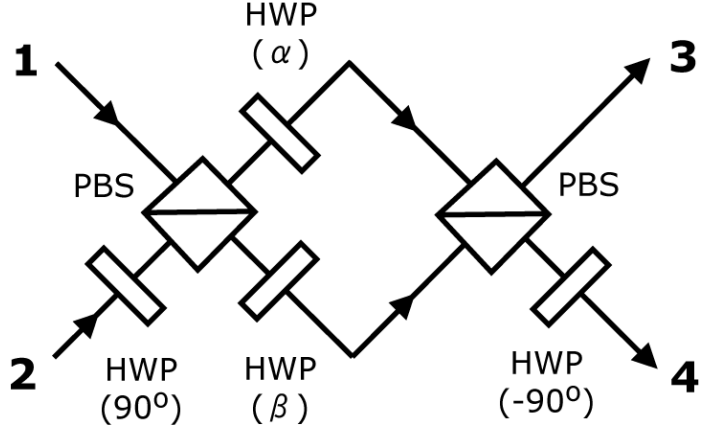}
\caption{ Implementation of the polarization sensitive variable
beam splitter with reflectivity probabilities $R_{\text{V}}$ and
$R_{\text{H}}$ using polarization-optical elements. The setup is
constructed by two PBS's and four HWP's. The first (last) HWP
rotate the polarization state 90(-90) degree. The two middle
HWP's inside the interferometer rotate the polarization state
arbitrary angles $\alpha $ and $\beta $. $R_{\text{V}}$ and
$R_{\text{H}}$ are determined by the rotation angles $\alpha $
and $\beta $, and become $R_{\text{V}}=\cos ^{2}\alpha $ and
$R_{\text{H}}=\cos^{2}\beta $.} \label{fig3}
\end{figure}

One could, in principle, build a custom optical beam splitter
(i.e. dielectric filters) with the desired reflectivities to
satisfy (\ref{rv}) and (\ref{rh}) - in this case the scheme has
no phase sensitivity whatsoever. However, building such an optic
might have technology (and cost) constrains; here we describe a
simple method to construct such a polarization-sensitive beam
splitter using standard polarization optical components with the
tradeoff of accepting some phase sensitivity. This polarization
sensitive variable beam splitter can be constructed constructed
by two PBS's and four HWP's that is shown in Fig. \ref{fig3}. The
first (last) HWP that is placed on mode 2 (4) rotate the
polarization state 90 (-90) degrees. The middle two HWP's in the
interferometer rotate the polarization state arbitrary angles
$\alpha $ and $\beta $. The two input modes, 1 and 2, undergo a
unitary transformation into two output modes, 3 and 4. The
vertically-polarized photon is described by
{\setcounter{enumi}{%
\value{equation}} \addtocounter{enumi}{1} \setcounter{equation}{0} %
\renewcommand{\theequation}{\theenumi\alph{equation}}
\begin{eqnarray}
\widehat{a}_{\text{V1}}\rightarrow \cos \alpha \,\,\widehat{a}_{\text{V3}}+\sin \alpha \,\, \widehat{a}_{\text{V4}},  \label{hm1} \\
\widehat{a}_{\text{V2}}\rightarrow -\sin \alpha
\,\,\widehat{a}_{\text{V3}}+\cos \alpha
\,\,\widehat{a}_{\text{V4}},  \label{hm2}
\end{eqnarray}
\setcounter{equation}{\value{enumi}} }and the
horizontally-polarized photon is described by
{\setcounter{enumi}{%
\value{equation}} \addtocounter{enumi}{1} \setcounter{equation}{0} %
\renewcommand{\theequation}{\theenumi\alph{equation}}
\begin{eqnarray}
\widehat{a}_{\text{H1}}\rightarrow \cos \beta \,\,\widehat{a}_{\text{H3}}+\sin \beta \,\widehat{a}_{\text{H4}},  \label{vm1} \\
\widehat{a}_{\text{H2}}\rightarrow -\sin \beta
\,\,\widehat{a}_{\text{H3}}+\cos \beta
\,\,\widehat{a}_{\text{H4}}.  \label{vm2}
\end{eqnarray}
\setcounter{equation}{\value{enumi}} }$R_{\text{V}}$ and
$R_{\text{H}}$ are determined by the rotation angles $\alpha $
and $\beta $, and become $R_{\text{V}}=\cos ^{2}\alpha $ and
$R_{\text{H}}=\cos ^{2}\beta $. When the angles are set $\alpha
\approx$ 29.5 degrees and $\beta \approx $ 61.6 degrees, required
reflectivities (\ref{rv}) and (\ref{rh}) is satisfied.

This theoretical paper details how our recently demonstrated
nonlinear sign shift gate can be used to construct the requisite
two-qubit conditional phase gate for scalable linear optics
quantum computation\cite{Sanaka2}. The typical 50/50 BS, where
$R_{\text{V}}=R_{\text{H}}=1/2$, is used for the experiment. The
presented scheme is the readily feasible extension of the
experiment by simply using two NS gates and changing the
reflectivity of the BS. The scheme herein is also designed to
eliminate all or most of the phase sensitivity inherent in other
linear optics gates to simplify the experimental implementation.
We have also described a simple method for building polarization
sensitive variable beam splitters - such an object may have other
uses in quantum optics like optimal phase covariant cloning of
photonic qubits\cite{Fiurasek}. This work and the recent results
bring linear optics quantum computation and information processing
closer to a reality.

This work was supported by the Austrian Science Foundation (FWF),
project numbers M666 and SFB 015 P06, NSERC, and the European
Commission, contract number IST-2001-38864.

\newpage

\end{document}